\documentclass[11pt]{article}
\usepackage{amsmath,amssymb,amsfonts,bbm}
\usepackage{amsmath,amssymb,amsfonts,bbm}
\usepackage{multirow}
\usepackage[table,xcdraw]{xcolor}
\usepackage{graphicx}
\usepackage{caption}
\usepackage{subcaption}
\usepackage[export]{adjustbox}

\newcommand{\be}{\begin{equation}}
\newcommand{\ee}{\end{equation}}
\newcommand{\bea}{\begin{eqnarray}}
\newcommand{\eea}{\end{eqnarray}}
\newcommand{\ba}{\begin{array}}
\newcommand{\ea}{\end{array}}

\newcommand{\htwo}{h_{2,1}}
\newcommand{\M}{\mathcal{M}}
\newcommand{\N}{\mathcal{N}}

\newcommand{\K}{\mathcal{K}}

\long\def\symbolfootnote[#1]#2{\begingroup%
\def\thefootnote{\fnsymbol{footnote}}\footnote[#1]{#2}\endgroup}

\begin{document}

\thispagestyle{empty}\vspace{40pt}

\hfill{}

\vspace{128pt}

\begin{center}
    \textbf{\Large BPS brane cosmology in $\N=2$ supergravity}\\
    \vspace{40pt}

    Moataz H. Emam\symbolfootnote[1]{\tt moataz.emam@cortland.edu}

    \vspace{12pt}   \textit{Department of Physics}\\
                    \textit{SUNY College at Cortland}\\
                    \textit{Cortland, New York 13045, USA}\\
\end{center}

\vspace{40pt}

\begin{abstract}

We study the embedding of flat BPS 3-branes in five dimensional $\N=2$ supergravity theory. We derive the branes' dynamical equations as well as general expressions for the hypermultiplet fields then focus on a single brane and study its time evolution. It is shown that the brane's Hubble parameter correlates with the moduli of the underlying manifold's complex structure. For certain particular solutions, the moduli seem to exhibit an instability; being large valued at early times then rapidly decaying to either zero or some convergent constant value. The possibility of extending these results to the cosmology of our universe is implied and briefly discussed. Our results are in line with the production and decay of heavy moduli in the early universe, as is currently believed in the literature.

\end{abstract}

\newpage



\vspace{15pt}

\pagebreak

\section{Introduction}

The study of higher dimensional branes is an important part of the ongoing quest to understand the structure of superstring theory. It is generally motivated by their possible roles in understanding the string-theoretic origins of entropy, duality symmetries, the AdS/CFT correspondence etc \cite{Aharony:1999ti}. Particularly useful is how branes relate to models of dimensional reduction and/or large extra dimensions. It has long been the hope that a fully quantum mechanical theory of branes will lay the groundwork for a full exposition of nonperturbative string theory \cite{Duff:1999rk}. It is within this general view of `beyond the standard model' physics that much work has been done to classify all possible fully or partially supersymmetric brane configurations. This is usually performed within the boundaries of supergravity theory which, while admittedly a classical low energy version of the full string theory, is nonperturbative and hence expose properties of branes that would be difficult to explore in a perturbative approach. Furthermore, the possible interpretation of our universe as a 3-brane imbedded in a higher dimensional bulk adds even more interest to brane-theory and has understandably generated a lot of research in recent years, starting with the seminal work by Randall and Sundrum \cite{Randall:1999ee}. Since then, various models of `brane-cosmology' have been proposed \cite{Brax:2003fv, Maartens:2010ar, Roane:2007zz}. Most of which present studies of expanding (supersymmetric or non-supersymmetric) brane-universes via various stages of their evolution: inflation, re-heating, slow acceleration etc as well as possible `explanations' for the big bang itself (\emph{e.g.} \cite{Flanagan:1999cu, Binetruy:1999hy, Saaidi:2010jw, Saaidi:2012ri, Lidsey:2000mt, Choudhury:2012ib, Maia:2008yya, Okada:2014eva, Cordero:2011zz, Capistrano:2011zz, Amarilla:2009rs, Carmeli:1900zzc, Koyama:2007rx, Antoniadis:2007hp, McFadden:2005mq, Garriga:2001qt, Rasanen:2001hf, Khoury:2001wf, Falkowski:2000er}). It seems that the universe has always been in some form of accelerating expansion. While the current stage of slow acceleration is attributed to the cosmological constant/vacuum energy, the more expansive abrupt inflationary acceleration is explained by assuming the existence of the so-called `inflaton'; a scalar field active in the early universe \cite{astro-ph/9805201, Linde:1994yf}. Various models within the string theory landscape have been presented to explain either the current value of the cosmological constant or the inflaton field (e.g. \cite{AvilanV.:2010ri, 1203.0307, Park:2013nu, Kallosh:2010xz, Gong:2006be}). But there does not seem to be any studies that attempt to explore the history of the universe from inflation through the current phase, in other words no single model exists to explain why the universe passes through accelerating phases with various magnitudes. In this paper, we study a 3-brane embedded in five dimensional ungauged $\N=2$ supergravity with bulk hypermultiplets and find that the moduli of the complex structure of the underlying Calabi-Yau (CY) space act as a possible source for the various cosmological stages of said brane. The brane is vacuous, \emph{i.e.} devoid of all matter and radiation. We show that its spatial scale, described by a Robertson-Walker like scale factor $a\left(t\right)$, is dependent on the norm of the moduli of the CY complex structure, hence by considering various forms for $a\left(t\right)$ one can calculate, in reverse, the behavior of the moduli. We consider a generalized expansion model, where the brane is allowed to go through an inflationary phase, followed by a slow accelerative expansion, and find that these stages correlate to a behavior of the moduli that seems to suggest a high degree of instability. The norm of the moduli begins at a very high value then rapidly decays (synchronous with inflation) tending to a constant value at infinite times. From a cosmological perspective this is in agreement with the conjecture of the early production of heavy moduli and subsequent decay, most likely into gravitinos, as required by the phenomenology of the early universe. The configuration studied instantaneously satisfies the Bogomol'nyi-Prasad-Sommerfield (BPS) condition and breaks half of the supersymmetries under certain constraints which we also derive. There also seems to be a lot of freedom as to the exact form of the hypermultiplet fields, depending on the explicit form of the moduli as well as a bulk harmonic function. In addition to being an interesting result from the point of view of pure brane theory, one hopes that it will lay the ground work for further investigation of the possible effect of the complex structure moduli \cite{Hayashi:2014aua} on the evolution of our universe.

\section{$D=5$ $\N=2$ supergravity with hypermultiplets} \label{theory}

The dimensional reduction of $D=11$ supergravity theory over a Calabi-Yau 3-fold $\M$ with nontrivial complex structure moduli yields an $\N=2$ supergravity theory in $D=5$ with a set of scalar fields and their supersymmetric partners all together known as the \emph{hypermultiplets} (see \cite{Emam:2010kt} for a review and additional references). It should be noted that the other matter sector in the theory; the vector multiplets, trivially decouples from the hypermultiplets and can simply be set to zero, as we do here. The hypermultiplets are partially comprised of the \emph{universal hypermultiplet} $\left(\varphi, \sigma, \zeta^0, \tilde \zeta_0\right)$; so called because it appears irrespective of the detailed structure of the sub-manifold. The field $\varphi$ is known as the universal axion, and is magnetically dual to a three-form gauge field and the dilaton $\sigma$ is proportional to the natural logarithm of the volume of $\M$. The rest of the hypermultiplets are $\left(z^i, z^{\bar i}, \zeta^i, \tilde \zeta_i: i=1,\ldots, \htwo\right)$, where the $z$'s are identified with the complex structure moduli of $\M$, and $\htwo$ is the Hodge number determining the dimensions of the manifold of the Calabi-Yau's complex structure moduli, $\M_C$. The `bar' over an index denotes complex conjugation. The fields $\left(\zeta^I, \tilde\zeta_I: I=0,\ldots,\htwo\right)$ are known as the axions and arise as a result of the $D=11$ Chern-Simons term. The supersymmetric partners known as the hyperini complete the hypermultiplets.

The theory has a very rich structure that arises from the intricate topology of $\M$. Of particular usefulness is its symplectic covariance. Specifically, the axions $\left(\zeta^I, \tilde\zeta_I\right)$ can be defined as components of the symplectic vector
\be\label{DefOfSympVect}
   \left| \Xi  \right\rangle  = \left( {\begin{array}{*{20}c}
   {\,\,\,\,\,\zeta ^I }  \\
   -{\tilde \zeta _I }  \\
    \end{array}} \right),
\ee
such that the symplectic scalar product is defined by, for example,
\be
    \left\langle {{\Xi }}
 \mathrel{\left | {\vphantom {{\Xi } d\Xi }}
 \right. \kern-\nulldelimiterspace}
 {d\Xi } \right\rangle   = \zeta^I d\tilde \zeta_I  - \tilde \zeta_I
 d\zeta^I,\label{DefOfSympScalarProduct}
\ee
where $d$ is the spacetime exterior derivative $\left(d=dx^\mu\partial_\mu:\mu=0,\ldots,4\right)$. A `rotation' in symplectic space is defined by the matrix element
\bea
 \left\langle {\partial _\mu \Xi } \right|{\bf\Lambda} \left| {\partial ^\mu \Xi } \right\rangle \star \mathbf{1} &=& \left\langle {d\Xi } \right|\mathop {\bf\Lambda} \limits_ \wedge  \left| {\star d\Xi } \right\rangle  \nonumber\\
  &=& 2\left\langle {{d\Xi }}
 \mathrel{\left | {\vphantom {{d\Xi } V}}
 \right. \kern-\nulldelimiterspace}
 {V} \right\rangle \mathop {}\limits_ \wedge  \left\langle {{\bar V}}
 \mathrel{\left | {\vphantom {{\bar V} {\star d\Xi }}}
 \right. \kern-\nulldelimiterspace}
 {{\star d\Xi }} \right\rangle  + 2G^{i\bar j} \left\langle {{d\Xi }}
 \mathrel{\left | {\vphantom {{d\Xi } {U_{\bar j} }}}
 \right. \kern-\nulldelimiterspace}
 {{U_{\bar j} }} \right\rangle \mathop {}\limits_ \wedge  \left\langle {{U_i }}
 \mathrel{\left | {\vphantom {{U_i } {\star d\Xi }}}
 \right. \kern-\nulldelimiterspace}
 {{\star d\Xi }} \right\rangle  - i\left\langle {d\Xi } \right.\mathop |\limits_ \wedge  \left. {\star d\Xi } \right\rangle,\label{DefOfRotInSympSpace}
\eea
where $\star$ is the $D=5$ Hodge duality operator, and $G_{i\bar j}$ is a special K\"{a}hler metric on $\M_C$. The symplectic basis vectors $\left| V \right\rangle $, $\left| {U_i } \right\rangle $ and their complex conjugates are defined by
\be
    \left| V \right\rangle  = e^{\frac{\K}{2}} \left( {\begin{array}{*{20}c}
   {Z^I }  \\
   {F_I }  \\
    \end{array}} \right),\,\,\,\,\,\,\,\,\,\,\,\,\,\,\,\left| {\bar V} \right\rangle  = e^{\frac{\K}{2}} \left( {\begin{array}{*{20}c}
   {\bar Z^I }  \\
   {\bar F_I }  \\
    \end{array}} \right)\label{DefOfVAndVBar}
\ee
where $\K$ is the K\"{a}hler potential on $\M_C$, $\left( {Z,F} \right)$ are the periods of the Calabi-Yau's holomorphic volume form, and
\bea
    \left| {U_i } \right\rangle  &=& \left| \nabla _i V
    \right\rangle=\left|\left[ {\partial _i  + \frac{1}{2}\left( {\partial _i \K} \right)} \right] V \right\rangle \nonumber\\
    \left| {U_{\bar i} } \right\rangle  &=& \left|\nabla _{\bar i}  {\bar V} \right\rangle=\left|\left[ {\partial _{\bar i}  + \frac{1}{2}\left( {\partial _{\bar i} \K} \right)} \right] {\bar V}
    \right\rangle\label{DefOfUAndUBar}
\eea
where the derivatives are with respect to the moduli $\left(z^i, z^{\bar i}\right)$. These vectors satisfy the following conditions:
\bea
    \left\langle {{\bar V}}
     \mathrel{\left | {\vphantom {{\bar V} V}}
     \right. \kern-\nulldelimiterspace}
     {V} \right\rangle   &=& i\nonumber\\
    \left|\nabla _i  {\bar V} \right\rangle  &=& \left|\nabla _{\bar i}  V \right\rangle =0\nonumber\\
    \left\langle {{U_i }}
    \mathrel{\left | {\vphantom {{U_i } {U_j }}}
    \right. \kern-\nulldelimiterspace}
    {{U_j }} \right\rangle  &=& \left\langle {{U_{\bar i} }}
    \mathrel{\left | {\vphantom {{U_{\bar i} } {U_{\bar j} }}}
    \right. \kern-\nulldelimiterspace}
    {{U_{\bar j} }} \right\rangle    =0\nonumber\\
    \left\langle {\bar V}
    \mathrel{\left | {\vphantom {\bar V {U_i }}}
    \right. \kern-\nulldelimiterspace}
    {{U_i }} \right\rangle  &=& \left\langle {V}
    \mathrel{\left | {\vphantom {V {U_{\bar i} }}}
    \right. \kern-\nulldelimiterspace}
    {{U_{\bar i} }} \right\rangle  = \left\langle { V}
    \mathrel{\left | {\vphantom { V {U_i }}}
    \right. \kern-\nulldelimiterspace}
    {{U_i }} \right\rangle=\left\langle {\bar V}
    \mathrel{\left | {\vphantom {\bar V {U_{\bar i} }}}
    \right. \kern-\nulldelimiterspace}
    {{U_{\bar i} }} \right\rangle= 0,\nonumber\\
    \left|\nabla _{\bar j}  {U_i } \right\rangle  &=& G_{i\bar j} \left| V \right\rangle ,\quad \quad \left|\nabla _i  {U_{\bar j} } \right\rangle  = G_{i\bar j} \left| {\bar V}
    \right\rangle,\nonumber\\
    G_{i\bar j}&=& \left( {\partial _i \partial _{\bar j} \K} \right)=- i    \left\langle {{U_i }}
    \mathrel{\left | {\vphantom {{U_i } {U_{\bar j} }}}
    \right. \kern-\nulldelimiterspace}
    {{U_{\bar j} }} \right\rangle.
\eea

The origin of these identities lies in special K\"{a}hler geometry. In our previous work \cite{Emam:2009xj}, we derived the following useful formulae:
\bea
    dG_{i\bar j}  &=& G_{k\bar j} \Gamma _{ri}^k dz^r  + G_{i\bar k} \Gamma _{\bar r\bar j}^{\bar k} dz^{\bar r}  \nonumber\\
    dG^{i\bar j}  &=&  - G^{p\bar j} \Gamma _{rp}^i dz^r  - G^{i\bar p} \Gamma _{\bar r\bar p}^{\bar j} dz^{\bar r}  \nonumber\\
    \left| {dV} \right\rangle  &=& dz^i \left| {U_i } \right\rangle  - i\mathfrak{Im} \left[ {\left( {\partial_i  \K} \right)dz^i} \right]\left| V \right\rangle \nonumber \\
    \left| {d\bar V} \right\rangle  &=& dz^{\bar i} \left| {U_{\bar i} } \right\rangle  + i\mathfrak{Im} \left[ {\left( {\partial_i  \K} \right)dz^i} \right]\left| {\bar V} \right\rangle \nonumber
\\
    \left| {dU_i } \right\rangle  &=& G_{i\bar j} dz^{\bar j} \left| V \right\rangle  + \Gamma _{ik}^r dz^k \left| {U_r } \right\rangle+G^{j\bar l} C_{ijk} dz^k \left| {U_{\bar l} } \right\rangle - i\mathfrak{Im} \left[ {\left( {\partial_i  \K} \right)dz^i} \right]\left| {U_i } \right\rangle \nonumber \\
    \left| {dU_{\bar i} } \right\rangle  &=& G_{j\bar i} dz^j \left| {\bar V} \right\rangle + \Gamma _{\bar i\bar k}^{\bar r} dz^{\bar k} \left| {U_{\bar r} } \right\rangle + G^{l\bar j} C_{\bar i\bar j\bar k} dz^{\bar k} \left| {U_l } \right\rangle + i\mathfrak{Im} \left[ {\left( {\partial_i  \K} \right)dz^i} \right]\left| {U_{\bar i} } \right\rangle\nonumber
\\
     {\bf \Lambda } &=& 2\left| V \right\rangle \left\langle {\bar V} \right| + 2G^{i\bar j} \left| {U_{\bar j} } \right\rangle \left\langle {U_i } \right|
    -i\nonumber\\
    {\bf \Lambda }^{-1} &=& -2\left| V \right\rangle \left\langle {\bar V} \right| - 2G^{i\bar j} \left| {U_{\bar j} } \right\rangle \left\langle {U_i } \right|
    +i\nonumber\\
        \partial_i {\bf \Lambda } &=& 2\left| {U_i } \right\rangle \left\langle {\bar V} \right|+2\left| {\bar V} \right\rangle \left\langle {U_i } \right| + 2G^{j\bar r} G^{k\bar p} C_{ijk} \left| {U_{\bar r} } \right\rangle \left\langle {U_{\bar p} } \right|.
\eea

The quantities $C_{ijk}$ are the components of the totally symmetric tensor that appears in the curvature tensor of $\M_C$. In this language, the bosonic part of the action is given by:
\bea
    S_5  &=& \int\limits_5 {\left[ {R\star \mathbf{1} - \frac{1}{2}d\sigma \wedge\star d\sigma  - G_{i\bar j} dz^i \wedge\star dz^{\bar j} } \right.}  + e^\sigma   \left\langle {d\Xi } \right|\mathop {\bf\Lambda} \limits_ \wedge  \left| {\star d\Xi } \right\rangle\nonumber\\
    & &\left. {\quad\quad\quad\quad\quad\quad\quad\quad\quad\quad\quad\quad\quad - \frac{1}{2} e^{2\sigma } \left[ {d\varphi + \left\langle {\Xi } \mathrel{\left | {\vphantom {\Xi  {d\Xi }}} \right. \kern-\nulldelimiterspace} {{d\Xi }}    \right\rangle} \right] \wedge \star\left[ {d\varphi + \left\langle {\Xi } \mathrel{\left | {\vphantom {\Xi  {d\Xi }}} \right. \kern-\nulldelimiterspace} {{d\Xi }}    \right\rangle} \right] } \right].\label{action}
\eea

The variation of the action yields the following field equations for $\sigma$, $\left(z^i,z^{\bar i}\right)$, $\left| \Xi  \right\rangle$ and $\varphi$ respectively:
\bea
    \left( {\Delta \sigma } \right)\star \mathbf{1} + e^\sigma   \left\langle {d\Xi } \right|\mathop {\bf\Lambda} \limits_ \wedge  \left| {\star d\Xi } \right\rangle -   e^{2\sigma }\left[ {d\varphi + \left\langle {\Xi } \mathrel{\left | {\vphantom {\Xi  {d\Xi }}} \right. \kern-\nulldelimiterspace} {{d\Xi }}    \right\rangle} \right]\wedge\star\left[ {d\varphi + \left\langle {\Xi } \mathrel{\left | {\vphantom {\Xi  {d\Xi }}} \right. \kern-\nulldelimiterspace} {{d\Xi }}    \right\rangle} \right] &=& 0\label{DilatonEOM}\\
    \left( {\Delta z^i } \right)\star \mathbf{1} + \Gamma _{jk}^i dz^j  \wedge \star dz^k  + \frac{1}{2}e^\sigma  G^{i\bar j}  {\partial _{\bar j} \left\langle {d\Xi } \right|\mathop {\bf\Lambda} \limits_ \wedge  \left| {\star d\Xi } \right\rangle} &=& 0 \nonumber\\
    \left( {\Delta z^{\bar i} } \right)\star \mathbf{1} + \Gamma _{\bar j\bar k}^{\bar i} dz^{\bar j}  \wedge \star dz^{\bar k}  + \frac{1}{2}e^\sigma  G^{\bar ij}  {\partial _j \left\langle {d\Xi } \right|\mathop {\bf\Lambda} \limits_ \wedge  \left| {\star d\Xi } \right\rangle}  &=& 0\label{ZZBarEOM} \\
    d^{\dag} \left\{ {e^\sigma  \left| {{\bf\Lambda} d\Xi } \right\rangle  - e^{2\sigma } \left[ {d\varphi + \left\langle {\Xi }
    \mathrel{\left | {\vphantom {\Xi  {d\Xi }}}\right. \kern-\nulldelimiterspace} {{d\Xi }} \right\rangle } \right]\left| \Xi  \right\rangle } \right\} &=& 0\label{AxionsEOM}\\
    d^{\dag} \left[ {e^{2\sigma } d\varphi + e^{2\sigma } \left\langle {\Xi } \mathrel{\left | {\vphantom {\Xi  {d\Xi }}} \right. \kern-\nulldelimiterspace} {{d\Xi }}    \right\rangle} \right] &=&    0\label{aEOM}
\eea
where $d^\dagger$ is the $D=5$ adjoint exterior derivative, $\Delta$ is the Laplace-de Rahm operator and $\Gamma _{jk}^i$ is a connection on $\M_C$.

The full action is symmetric under the following SUSY transformations:
\bea
 \delta _\epsilon  \psi ^1  &=& D \epsilon _1  + \frac{1}{4}\left\{ {i {e^{\sigma } \left[ {d\varphi + \left\langle {\Xi }
 \mathrel{\left | {\vphantom {\Xi  {d\Xi }}}
 \right. \kern-\nulldelimiterspace} {{d\Xi }} \right\rangle } \right]}- Y} \right\}\epsilon _1  - e^{\frac{\sigma }{2}} \left\langle {{\bar V}}
 \mathrel{\left | {\vphantom {{\bar V} {d\Xi }}} \right. \kern-\nulldelimiterspace} {{d\Xi }} \right\rangle\epsilon _2  \nonumber\\
 \delta _\epsilon  \psi ^2  &=& D \epsilon _2  - \frac{1}{4}\left\{ {i {e^{\sigma } \left[ {d\varphi + \left\langle {\Xi }
 \mathrel{\left | {\vphantom {\Xi  {d\Xi }}} \right. \kern-\nulldelimiterspace}
 {{d\Xi }} \right\rangle } \right]}- Y} \right\}\epsilon _2  + e^{\frac{\sigma }{2}} \left\langle {V}
 \mathrel{\left | {\vphantom {V {d\Xi }}} \right. \kern-\nulldelimiterspace} {{d\Xi }} \right\rangle \epsilon _1  \label{SUSYGraviton}
\\
  \delta _\epsilon  \xi _1^0  &=& e^{\frac{\sigma }{2}} \left\langle {V}
    \mathrel{\left | {\vphantom {V {\partial _\mu  \Xi }}} \right. \kern-\nulldelimiterspace} {{\partial _\mu  \Xi }} \right\rangle  \Gamma ^\mu  \epsilon _1  - \left\{ {\frac{1}{2}\left( {\partial _\mu  \sigma } \right) - \frac{i}{2} e^{\sigma } \left[ {\left(\partial _\mu \varphi\right) + \left\langle {\Xi }
    \mathrel{\left | {\vphantom {\Xi  {\partial _\mu \Xi }}} \right. \kern-\nulldelimiterspace}
    {{\partial _\mu \Xi }} \right\rangle } \right]} \right\}\Gamma ^\mu  \epsilon _2  \nonumber\\
     \delta _\epsilon  \xi _2^0  &=& e^{\frac{\sigma }{2}} \left\langle {{\bar V}}
    \mathrel{\left | {\vphantom {{\bar V} {\partial _\mu  \Xi }}} \right. \kern-\nulldelimiterspace} {{\partial _\mu  \Xi }} \right\rangle \Gamma ^\mu  \epsilon _2  + \left\{ {\frac{1}{2}\left( {\partial _\mu  \sigma } \right) + \frac{i}{2} e^{\sigma } \left[ {\left(\partial _\mu \varphi\right) + \left\langle {\Xi }
    \mathrel{\left | {\vphantom {\Xi  {\partial _\mu \Xi }}} \right. \kern-\nulldelimiterspace}
    {{\partial _\mu \Xi }} \right\rangle } \right]} \right\}\Gamma ^\mu  \epsilon
     _1\label{SUSYHyperon1}
\\
     \delta _\epsilon  \xi _1^{\hat i}  &=& e^{\frac{\sigma }{2}} e^{\hat ij} \left\langle {{U_j }}
    \mathrel{\left | {\vphantom {{U_j } {\partial _\mu  \Xi }}} \right. \kern-\nulldelimiterspace} {{\partial _\mu  \Xi }} \right\rangle \Gamma ^\mu  \epsilon _1  - e_{\,\,\,\bar j}^{\hat i} \left( {\partial _\mu  z^{\bar j} } \right)\Gamma ^\mu  \epsilon _2  \nonumber\\
     \delta _\epsilon  \xi _2^{\hat i}  &=& e^{\frac{\sigma }{2}} e^{\hat i\bar j} \left\langle {{U_{\bar j} }}
    \mathrel{\left | {\vphantom {{U_{\bar j} } {\partial _\mu  \Xi }}} \right. \kern-\nulldelimiterspace} {{\partial _\mu  \Xi }} \right\rangle \Gamma ^\mu  \epsilon _2  + e_{\,\,\,j}^{\hat i} \left( {\partial _\mu  z^j } \right)\Gamma ^\mu  \epsilon    _1,\label{SUSYHyperon2}
\eea
where $\left(\psi ^1, \psi ^2\right)$ are the two gravitini and $\left(\xi _1^I, \xi _2^I\right)$ are the hyperini. The quantity $Y$ is defined by:
\begin{equation}
    Y   = \frac{{\bar Z^I N_{IJ}  {d  Z^J }  -
    Z^I N_{IJ}  {d  \bar Z^J } }}{{\bar Z^I N_{IJ} Z^J
    }},\label{DefOfY}
\end{equation}
where $N_{IJ}  = \mathfrak{Im} \left({\partial_IF_J } \right)$. The $e$'s are the beins of the special K\"{a}hler metric $G_{i\bar j}$, the $\epsilon$'s are the five-dimensional $\N=2$ SUSY spinors and the $\Gamma$'s are the usual Dirac matrices. The covariant derivative $D$ is given by $D=dx^\mu\left( \partial _\mu   + \frac{1}{4}\omega _\mu^{\,\,\,\,\hat \mu\hat \nu} \Gamma _{\hat \mu\hat \nu}\right)\label{DefOfCovDerivative}$ as usual, where the $\omega$'s are the spin connections and the hatted indices are frame indices in a flat tangent space. Finally, the stress tensor is:
\bea
T_{\mu \nu }  &=& -\frac{1}{2}\left( {\partial _\mu  \sigma } \right)\left( {\partial _\nu  \sigma } \right) + \frac{1}{4}g_{\mu \nu } \left( {\partial _\alpha  \sigma } \right)\left( {\partial ^\alpha  \sigma } \right)
 + e^\sigma  \left\langle {\partial _\mu \Xi } \right|{\bf\Lambda} \left| {\partial _\nu \Xi } \right\rangle - \frac{1}{2}e^{\sigma } g_{\mu \nu }   \left\langle {\partial _\alpha \Xi } \right|{\bf\Lambda} \left| {\partial ^\alpha \Xi } \right\rangle \nonumber\\
  & &  - \frac{1}{2}e^{2\sigma } \left[ {\left( {\partial _\mu  \varphi} \right) + \left\langle {\Xi }
 \mathrel{\left | {\vphantom {\Xi  {\partial _\mu  \Xi }}}
 \right. \kern-\nulldelimiterspace}
 {{\partial _\mu  \Xi }} \right\rangle } \right]\left[ {\left( {\partial _\nu  \varphi} \right) + \left\langle {\Xi }
 \mathrel{\left | {\vphantom {\Xi  {\partial _\nu  \Xi }}}
 \right. \kern-\nulldelimiterspace}
 {{\partial _\nu  \Xi }} \right\rangle } \right]
  +  \frac{1}{4}e^{2\sigma } g_{\mu \nu } \left[ {\left( {\partial _\alpha  \varphi} \right) + \left\langle {\Xi }
 \mathrel{\left | {\vphantom {\Xi  {\partial _\alpha  \Xi }}}
 \right. \kern-\nulldelimiterspace}
 {{\partial _\alpha  \Xi }} \right\rangle } \right]\left[ {\left( {\partial ^\alpha  \varphi} \right) + \left\langle {\Xi }
 \mathrel{\left | {\vphantom {\Xi  {\partial ^\alpha  \Xi }}}
 \right. \kern-\nulldelimiterspace}
 {{\partial ^\alpha  \Xi }} \right\rangle } \right]\nonumber\\
 & & - G_{i\bar j} \left( {\partial _\mu  z^i } \right)\left( {\partial _\nu  z^{\bar j} } \right) + \frac{1}{2}g_{\mu \nu } G_{i\bar j} \left( {\partial _\alpha  z^i } \right)\left( {\partial ^\alpha  z^{\bar j} } \right).\label{StressTensor}
\eea

\section{Brane dynamics and bulk fields configurations}\label{Braneanalysis}

We begin with a metric of the form
\be
    ds^2  =  - e^{2\alpha \left( {t,y} \right)} dt^2  + e^{2\beta \left( {t,y} \right)} \left( {dr^2  + r^2 d\Omega ^2 } \right) + e^{2\gamma \left( {t,y} \right)} dy^2\label{GeneralBrane}
\ee
where ${d\Omega ^2  = d\theta ^2  + \sin ^2 \left( \theta  \right)d\phi ^2 }$. This metric may be interpreted as representing a single 3-brane located at $y=0$ in the transverse space, it may also represent a stack of $N$ branes located at various values of $y=y_I$ $\left(I = 1, \ldots ,N \in \mathbb{Z}\right)$ where the warp functions $\alpha$, $\beta$, and $\gamma$ are rewritten such that $y\rightarrow \sum\limits_{I = 1}^N {\left| {y - y_I } \right|} $. Either way, we will eventually focus on the four dimensional $\left(t, r, \theta, \phi\right)$ dynamics, effectively evaluating the warp functions at a specific, but arbitrary, $y$ value. We also note that a metric of the form (\ref{GeneralBrane}) was shown in \cite{Kallosh:2001du} to be exactly the type needed for a consistent BPS cosmology. In addition, a model along similar lines was proposed and studied in \cite{Kabat:2001qt}.

The brane (or branes) is assumed completely vacuous, for the sake of simplicity, and as such it merely acts as a toy-model of a universe. To connect to a possible cosmological application a more realistic approach is needed, such as invoking the presence of the usual perfect fluid and a cosmological constant on the brane's surface, as well as possibly in the bulk (\emph{e.g.} \cite{Canestaro:2013xsa}). Based on this metric, the components of the Einstein tensor are
\bea
 G_{tt}  &=& 3\left( {\dot \beta ^2  + \dot \beta \dot \gamma } \right) - 3e^{2\left( {\alpha  - \gamma } \right)} \left( {\beta '' + 2\beta '^2  - \beta '\gamma '} \right) \nonumber\\
 G_{rr}  &=&  - e^{2\left( {\beta  - \alpha } \right)} \left[ {2\ddot \beta  + 3\dot \beta ^2  + \ddot \gamma  + \dot \gamma ^2  + 2\dot \beta \left( {\dot \gamma  - \dot \alpha } \right) - \dot \alpha \dot \gamma } \right] \nonumber\\
  & &+ e^{2\left( {\beta  - \gamma } \right)} \left[ {2\beta '' + 3\beta '^2  + \alpha '' + \alpha '^2  + 2\beta '\left( {\alpha ' - \gamma '} \right) - \alpha '\gamma '} \right] \nonumber\\
 G_{yy}  &=& 3\left( {\beta '^2  + \beta '\alpha '} \right) - 3e^{2\left( {\gamma  - \alpha } \right)} \left( {\ddot \beta  + 2\dot \beta ^2  - \dot \beta \dot \alpha } \right) \nonumber\\
 G_{yt}  &=& 3\left( {\dot \beta \alpha ' + \beta '\dot \gamma  - \dot \beta \beta ' - \dot \beta '} \right),
\eea
where a prime is a derivative with respect to $y$ and a dot is a derivative with respect to $t$. We are interested in bosonic configurations that preserve some supersymmetry, so the stress tensor (\ref{StressTensor}) can be considerably simplified by considering the vanishing of the supersymmetric variations (\ref{SUSYHyperon1}, \ref{SUSYHyperon2}), which may be rewritten in matrix form as follows
\be
    \left[ {\begin{array}{*{20}c}
   {2e^{\frac{\sigma }{2}} \left\langle {V}
    \mathrel{\left | {\vphantom {V {\partial _\mu  \Xi }}} \right. \kern-\nulldelimiterspace} {{\partial _\mu  \Xi }} \right\rangle  \Gamma ^\mu} &  {-\left\{  {\left( {\partial _\mu  \sigma } \right) - i e^{\sigma } \left[ {\left(\partial _\mu \varphi\right) + \left\langle {\Xi }
    \mathrel{\left | {\vphantom {\Xi  {\partial _\mu \Xi }}} \right. \kern-\nulldelimiterspace}
    {{\partial _\mu \Xi }} \right\rangle } \right]}  \right\}\Gamma ^\mu}  \\
   {}  & {}  \\
   {\left\{  {\left( {\partial _\mu  \sigma } \right) + i e^{\sigma } \left[ {\left(\partial _\mu \varphi\right) + \left\langle {\Xi }
    \mathrel{\left | {\vphantom {\Xi  {\partial _\mu \Xi }}} \right. \kern-\nulldelimiterspace}
    {{\partial _\mu \Xi }} \right\rangle } \right]}  \right\}\Gamma ^\nu} &  {2e^{\frac{\sigma }{2}} \left\langle {{\bar V}}
    \mathrel{\left | {\vphantom {{\bar V} {\partial _\nu  \Xi }}} \right. \kern-\nulldelimiterspace} {{\partial _\nu  \Xi }} \right\rangle \Gamma ^\nu}  \\
\end{array}} \right]\left( {\begin{array}{*{20}c}
   {\epsilon _1 }  \\
   {}  \\
   {\epsilon _2 }  \\
\end{array}} \right) = 0
\ee
\be
    \left[ {\begin{array}{*{20}c}
   {e^{\frac{\sigma }{2}} e^{\hat ij} \left\langle {{U_j }}
    \mathrel{\left | {\vphantom {{U_j } {\partial _\mu  \Xi }}} \right. \kern-\nulldelimiterspace} {{\partial _\mu  \Xi }} \right\rangle \Gamma ^\mu } & {} & {-e_{\,\,\,\bar j}^{\hat i} \left( {\partial _\mu  z^{\bar j} } \right)\Gamma ^\mu}  \\
   {} & {} & {}  \\
   {e_{\,\,\,k}^{\hat j} \left( {\partial _\nu  z^k } \right)\Gamma ^\nu} & {} & {e^{\frac{\sigma }{2}} e^{\hat j\bar k} \left\langle {{U_{\bar k} }}
    \mathrel{\left | {\vphantom {{U_{\bar j} } {\partial _\nu  \Xi }}} \right. \kern-\nulldelimiterspace} {{\partial _\mu  \Xi }} \right\rangle \Gamma ^\nu}  \\
\end{array}} \right]\left( {\begin{array}{*{20}c}
   {\epsilon _1 }  \\
   {}  \\
   {\epsilon _2 }  \\
\end{array}} \right) = 0.
\ee

The vanishing of the determinants gives the conditions:
\bea
 d\sigma  \wedge \star d\sigma  + e^{2\sigma } \left[ {d\varphi  + \left\langle {\Xi }
 \mathrel{\left | {\vphantom {\Xi  {d\Xi }}}
 \right. \kern-\nulldelimiterspace}
 {{d\Xi }} \right\rangle } \right] \wedge \star\left[ {d\varphi  + \left\langle {\Xi }
 \mathrel{\left | {\vphantom {\Xi  {d\Xi }}}
 \right. \kern-\nulldelimiterspace}
 {{d\Xi }} \right\rangle } \right]
 + 4e^\sigma  \left\langle {V}
 \mathrel{\left | {\vphantom {V {d\Xi }}}
 \right. \kern-\nulldelimiterspace}
 {{d\Xi }} \right\rangle  \wedge \left\langle {{\bar V}}
 \mathrel{\left | {\vphantom {{\bar V} {\star d\Xi }}}
 \right. \kern-\nulldelimiterspace}
 {{\star d\Xi }} \right\rangle  &=& 0 \nonumber\\
 G_{i\bar j} dz^i  \wedge \star dz^{\bar j}  + e^\sigma  G^{i\bar j} \left\langle {{U_i }}
 \mathrel{\left | {\vphantom {{U_i } {d\Xi }}}
 \right. \kern-\nulldelimiterspace}
 {{d\Xi }} \right\rangle  \wedge \left\langle {{U_{\bar j} }}
 \mathrel{\left | {\vphantom {{U_{\bar j} } {\star d\Xi }}}
 \right. \kern-\nulldelimiterspace}
 {{\star d\Xi }} \right\rangle  &=& 0.\label{FromSUSY}
\eea

Using this with (\ref{DefOfRotInSympSpace}) we find
\be
    e^\sigma  \left\langle {d\Xi } \right|\mathop {\bf\Lambda} \limits_ \wedge  \left| {\star d\Xi } \right\rangle  = \frac{1}{2}d\sigma  \wedge \star d\sigma  + \frac{1}{2}e^{2\sigma } \left[ {d\varphi  + \left\langle {\Xi }
 \mathrel{\left | {\vphantom {\Xi  {d\Xi }}}
 \right. \kern-\nulldelimiterspace}
 {{d\Xi }} \right\rangle } \right] \wedge \star\left[ {d\varphi  + \left\langle {\Xi }
 \mathrel{\left | {\vphantom {\Xi  {d\Xi }}}
 \right. \kern-\nulldelimiterspace}
 {{d\Xi }} \right\rangle } \right] + 2G_{i\bar j} dz^i  \wedge \star dz^{\bar j},\label{Rotation}
\ee
where we have used $\left\langle {d\Xi } \right.\mathop |\limits_ \wedge  \left. {\star d\Xi } \right\rangle  = 0$ as required by the reality of the axions. Using (\ref{Rotation}) in (\ref{StressTensor}) eliminates all terms involving $\sigma$, $\left| \Xi  \right\rangle$ and $\varphi$, leaving the dynamics to depend \emph{only} on the complex structure moduli $\left(z^i,z^{\bar i}\right)$:
\be
    T_{\mu \nu }  = G_{i\bar j} \left( {\partial _\mu  z^i } \right)\left( {\partial _\nu  z^{\bar j} } \right) - \frac{1}{2}g_{\mu \nu } G_{i\bar j} \left( {\partial _\alpha  z^i } \right)\left( {\partial ^\alpha  z^{\bar j} } \right).
\ee

The Einstein equations then yield
\bea
     \frac{1}{2}G_{i\bar j} \dot z^i \dot z^{\bar j}
 &=& - \left[ {2\ddot \beta  + 3\dot \beta ^2  + \ddot \gamma  + \dot \gamma ^2  + 2\dot \beta \left( {\dot \gamma  - \dot \alpha } \right) - \dot \alpha \dot \gamma } \right]\nonumber\\
 &=& - 3\left( {\ddot \beta  + 2\dot \beta ^2  - \dot \beta \dot \alpha } \right)\nonumber\\
 &=& 3\left( {\dot \beta ^2  + \dot \beta \dot \gamma } \right)\label{Gtt}\\
    \frac{1}{2}G_{i\bar j} {z^i}' {z^{\bar j}}'
 &=& -\left[{ 2\beta '' + 3\beta '^2  + \alpha '' + \alpha '^2  + 2\beta '\left( {\alpha ' - \gamma '} \right) - \alpha '\gamma ' }\right]\nonumber\\
 &=& - 3\left( {\beta '' + 2\beta '^2  - \beta '\gamma '} \right)\nonumber\\
 &=& 3\left( {\beta '^2  + \beta '\alpha '} \right)\label{G11}\\
   G_{i\bar j} {z^i}' \dot z^{\bar j}   &=& 3\left( {\dot \beta \alpha ' + \beta '\dot \gamma  - \dot \beta \beta ' - \dot \beta '} \right).\label{G21}
\eea

The right hand side equalities in (\ref{Gtt}, \ref{G11}) lead to
\bea
 \frac{{\ddot \gamma }}{{\dot \gamma }} &=& \frac{{\ddot \beta }}{{\dot \beta }}, \quad\quad
 \ddot \gamma  + \dot \gamma ^2  - \dot \alpha \dot \gamma  + 3\dot \beta \dot \gamma  = 0 \nonumber\\
 \frac{{\alpha ''}}{{\alpha '}} &=& \frac{{\beta ''}}{{\beta '}}, \quad\quad
 \alpha '' + \alpha '^2  - \alpha '\gamma ' + 3\beta '\alpha ' = 0.\label{AlphaBetaGamma}
\eea

Now some analysis of the field equations, independently of the metric, can be done as follows. Equations (\ref{AxionsEOM}) and (\ref{aEOM}) are already first integrals, which may be integrated to give
\be
     d\varphi +\left\langle {\Xi }
 \mathrel{\left | {\vphantom {\Xi  {d\Xi }}}
 \right. \kern-\nulldelimiterspace}
 {{d\Xi }} \right\rangle = n  e^{-2\sigma } dh,\label{universalaxionsolution}
\ee
where $h$ is harmonic in $\left(t,y\right)$, \emph{i.e.} satisfies $\Delta h = 0$, and $n  \in \mathbb{R}$. Similarly:
\be
    e^\sigma  \left| {{\bf\Lambda} d\Xi } \right\rangle  - n dh\left| \Xi  \right\rangle  = s \left| {dK} \right\rangle \,\,\,\,\,{\rm where}\,\,\,\,\,\left| {\Delta K} \right\rangle  = 0\,\,\,\,\,{\rm and}\,\,\,\,\,s  \in \mathbb{R}.\label{Axion-K}
\ee

To find an expression for the axions, we look again at the vanishing of the hyperini transformations (\ref{SUSYHyperon1}) and (\ref{SUSYHyperon2}) and make the simplifying assumption $\epsilon_1=\pm\epsilon_2$. This leads to:
\bea
    \left\langle {V}
 \mathrel{\left | {\vphantom {V {d\Xi }}}
 \right. \kern-\nulldelimiterspace}
 {{d\Xi }} \right\rangle &=& \frac{1}{2}e^{ - \frac{\sigma }{2}} d\sigma - \frac{{in}}{2}e^{ - \frac{3}{2}\sigma } dh\nonumber\\
    \left\langle {{\bar V}}
 \mathrel{\left | {\vphantom {{\bar V} {d\Xi }}}
 \right. \kern-\nulldelimiterspace}
 {{d\Xi }} \right\rangle &=& \frac{1}{2}e^{ - \frac{\sigma }{2}} d\sigma + \frac{{in}}{2}e^{ - \frac{3}{2}\sigma } dh
\nonumber\\
    \left\langle {{U_i }}
 \mathrel{\left | {\vphantom {{U_i } {d\Xi }}}
 \right. \kern-\nulldelimiterspace}
 {{d\Xi }} \right\rangle &=& e^{ - \frac{\sigma }{2}} G_{i\bar j} dz^{\bar j}\nonumber\\
    \left\langle {{U_{\bar j} }}
 \mathrel{\left | {\vphantom {{U_{\bar j} } {d\Xi }}}
 \right. \kern-\nulldelimiterspace}
 {{d\Xi }} \right\rangle &=& e^{ - \frac{\sigma }{2}} G_{i\bar j} dz^i.
\eea

These are the symplectic components of the full vector:
\be
 \left| {d\Xi } \right\rangle  =  e^{ - \frac{\sigma }{2}} \mathfrak{Re} \left[ {\left( {ne^{ - \sigma } dh- id\sigma} \right)\left| V \right\rangle   + 2i\left| {U_i } \right\rangle dz^i } \right].\label{dXi}
\ee

The reality condition $\overline{\left| {d\Xi } \right\rangle}  = \left| {d\Xi } \right\rangle $ as well as the Bianchi identity on the axions are trivially satisfied. Substituting (\ref{dXi}) in (\ref{Axion-K}), we get
\be
   \left| \Xi  \right\rangle dh = \frac{1}{n }e^{ - \frac{\sigma }{2}} \mathfrak{Re}\left[\left(d\sigma+in e^{-\sigma}dh\right)\left| V \right\rangle \right] + \frac{2}{n }e^{ - \frac{\sigma }{2}} \mathfrak{Re}\left[\left| U_i \right\rangle dz^i\right] - \frac{s}{n}e^{-\sigma}\left| {dK} \right\rangle .\label{42}
\ee

These general constraints on the axions are in fact as far as one can get here. The exact solutions depend on the moduli and the symplectic basis vectors, which in turn require knowledge of the underlying Calabi-Yau metric, which is of course unknown. On the other hand, the harmonic function $h$, which arises from $\Delta h =0$:
\be
    e^{\left( {\alpha  - \gamma } \right)} \left[ {h'' + \left( {\alpha ' + 3\beta ' - \gamma '} \right)h'} \right] = e^{\left( {\gamma  - \alpha } \right)} \left[ {\ddot h - \left( {\dot \alpha  - 3\dot \beta  - \dot \gamma } \right)\dot h} \right]\label{HarmonicH}
\ee
can be found, as we will see. Finally, the axions also depend on the dilaton. It can be shown that a simple ansatz for the dilaton (such as $\sigma\propto \ln h$, much used in the literature) is not satisfactory in this case and leads to trivial moduli. In fact, as we will see, the dilaton field equation turns out to be too complicated to solve generally. However, a certain special case solution can be written.

\section{The cosmology of a single brane}\label{cosmology}

The equations derived in the previous sections (specifically \ref{Gtt}, \ref{G11}, \ref{G21}, \ref{AlphaBetaGamma} and \ref{HarmonicH}) are the basic equations governing the dynamics of the multi-brane spacetime (\ref{GeneralBrane}). We may assume that the warp functions as well as $h$ are separable as follows:
\bea
    e^{\beta \left( {t,y} \right)}  &=& a\left( t \right)F\left( y \right)\nonumber\\
    e^{\gamma \left( {t,y} \right)}  &=& b\left( t \right)K\left( y \right)\nonumber\\
    e^{\alpha \left( {t,y} \right)}  &=& c\left( t \right)N\left( y \right)\nonumber\\
    h\left( {t,y} \right) &=& k\left( t \right)M\left( y \right).\label{Separation}
\eea

Our interest is the dynamics of a single brane out of an infinite number of possible 3-branes along $y$, so we will evaluate the functions $F\left( y \right)$, $K\left( y \right)$, $N\left( y \right)$ and $M\left( y \right)$ near the brane of interest and normalize the result to unity, \emph{i.e.} $F\left( 0 \right) = 1$ and so on, where the brane under study is located at $y=0$. The metric then becomes more Robertson-Walker like:
\be
    ds^2  =  - c^2 \left( t \right)dt^2  + a^2 \left( t \right) \left( {dr^2  + r^2 d\Omega ^2 } \right) + b^2 \left( t \right) dy^2,
\ee
and equations (\ref{Gtt}, \ref{G11}, \ref{G21}, \ref{AlphaBetaGamma} and \ref{HarmonicH}) simply reduce to:
\bea
 \frac{{\ddot b}}{{\dot b}} - \frac{{\dot b}}{b} &=& \frac{{\ddot a}}{{\dot a}} - \frac{{\dot a}}{a} \nonumber\\
 \left( {\frac{{\dot c}}{c}} \right)&=&\left( {\frac{{\ddot b}}{\dot b}} \right)+3\left( {\frac{{\dot a}}{a}} \right)\label{ScaleEquations1}\\
  G_{i\bar j} \dot z^i \dot z^{\bar j}&=& 6\left[ {\left( {\frac{{\dot a}}{a}} \right)^2  + \left( {\frac{{\dot a}}{a}} \right)\left( {\frac{{\dot b}}{b}} \right)} \right]\label{ScaleEquations2}\\
  G_{i\bar j} {z^i}' {z^{\bar j}}'&=&G_{i\bar j} {z^i}' \dot z^{\bar j} =0.\label{ScaleEquations3}\\
  \ddot k - \left[ {\left( {\frac{{\ddot a}}{\dot a}} \right)  - \left( {\frac{{\dot a}}{a}} \right)} \right]\dot k &=& p^2\left( {\frac{c}{b}} \right)^2 k, \quad\quad p \in \mathbb{R}.\label{ScaleEquations4}
\eea

Equations (\ref{ScaleEquations1}) can be exactly solved in terms of the brane's scale factor $a\left(t\right)$ as follows:
\bea
 b\left( t \right) &=& G_2 a^{G_1 }  \label{b_equation}\\
 c\left( t \right) &=& G_3 \dot aa^{2 + G_1 }\label{c_equation}
\eea
where $G_i \in \mathbb{R}$ are arbitrary integration constants. As noted earlier, complete solutions of the scalar fields of the hypermultiplets would require full knowledge of the structure of the underlying manifold $\M$. Some insight may be gained by solving equation (\ref{ScaleEquations4}) for $k$, since the harmonic function is the only connection between the metric's warp factors and the hypermultiplets in the bulk. Fortunately, this is easily done: For the special case of the vanishing of the separability constant $p=0$, we find
\be
    k\left( t \right) = G_5  + G_6 \ln a.
\ee

While for a general $p\ne 0$:
\be
    k\left( t \right)=G_5 I_0\left(\frac{p G_3 }{3 G_2} a^3   \right)+ G_6 K_0\left(\frac{p G_3 }{3 G_2}a^3\right),
\ee
where $I_0$ and $K_0$ are the modified Bessel functions of the first and second kinds respectively. In the previous section we found partially explicit forms for the axions, all dependent on $k$ and the moduli. If we direct our attention to the dilaton, its time dependence can be found from (\ref{DilatonEOM}):
\be
    \ddot \sigma  + \frac{1}{2}\dot \sigma ^2  = \frac{{n^2 }}{2}e^{ - 2\sigma } \dot k^2  - 12\left( {G_1  + 1} \right)\left( {\frac{{\dot a}}{a}} \right)^2,
\ee
which is unfortunately too complicated to solve in terms of an arbitrary $a$. We did find, however, one (rather trivial) solution for the case $p=0$, $G_1=-1$ and assuming $a\left(t\right) \propto e^{\omega t}$, where $\omega  \in \mathbb{R}$:
\be
    \sigma \left( t, 0 \right) = \ln \left[ {\frac{{G_7 }}{4}t^2  + \frac{{G_7 G_8 }}{2}t + \frac{{G_7 G_8^2 }}{4} + \frac{{n^2 \omega ^2 G_6^2 }}{{G_7 }}} \right].
\ee

Now, while again using the assumption $\epsilon_1=\pm\epsilon_2=\epsilon$, the vanishing of the gravitini equations (\ref{SUSYGraviton}) gives the following time dependent form for the near-brane spinors
\be
    \epsilon\left( t, 0 \right)  = e^{\frac{\sigma }{2} + \frac{3}{4}in\Omega k - \Upsilon } \hat \epsilon,
\ee
where $\hat \epsilon$ is an arbitrary constant spinor and the functions $\Upsilon$ and $\Omega$ are solutions of $\dot \Upsilon  = Y$ and $\frac{d}{{dt}}\left( {\Omega k} \right) = e^{ - \sigma } \dot k$.

From a cosmological perspective, the major result here follows from the moduli expression (\ref{ScaleEquations2}):
\be
     \left( {\frac{{\dot a}}{a}} \right)^2=\frac{G_{i\bar j} \dot z^i \dot z^{\bar j}}{6\left(G_1+1\right)}.\label{ModuliAsLambda}
\ee

This is a surprisingly simple Friedmann-type equation. It states that the brane-universe's Hubble parameter is proportional to the complex structure moduli (clearly this requires $G_1>-1$). Unless $G_{i\bar j} \dot z^i \dot z^{\bar j}$ is vanishing, one can correlate the value of the brane's acceleration at any time with the evolution of the space of complex structure moduli $\M_C$. Thinking in terms of the cosmology of our own universe, one could ask what form of the moduli should be assumed? From a phenomenological perspective, massive unstable moduli (of any type) must have been densely produced in the Big Bang itself, and the heavier they were the faster they must have decayed early on \cite{Dine:2006ii,Bodeker:2006ij}. While moduli stabilization is necessary for other reasons \cite{Dudas:2012wi}, most realistic scenarios involve rapidly decaying moduli. Using (\ref{ModuliAsLambda}), we find that any reasonable choice of $a$ would necessarily lead to the decay of the moduli, with varying degrees of instability, as expected. For example:
\bea
    a &=& t^\omega \quad\quad \rightarrow\quad\quad G_{i\bar j} \dot z^i \dot z^{\bar j}=\frac{{{\rm 6}\omega ^{\rm 2} (G_1  + 1)}}{{t^2 }}\nonumber\\
    a &=& e^{\omega t} \quad\quad \rightarrow\quad\quad G_{i\bar j} \dot z^i \dot z^{\bar j}={{\rm 6}\omega ^{\rm 2} (G_1  + 1)}\nonumber\\
    a &=& \ln\left(\omega t\right) \quad\quad \rightarrow\quad\quad G_{i\bar j} \dot z^i \dot z^{\bar j}=\frac{{6\left( {G_1  + 1} \right)}}{{t^2 \ln ^2 \left( {\omega t} \right)}},
\eea
and so on. More interestingly, one can choose a form of $a$ that represents different values of acceleration over $t\rightarrow 0$ and $t\rightarrow \infty$ epochs. This can be, for example:
\be
    a\left( t \right) = e^{{t \mathord{\left/ {\vphantom {t \omega }} \right. \kern-\nulldelimiterspace} \omega }}  - e^{ - \kappa t}, \quad\quad \omega ,\kappa  \in \mathbb{R} > 0\label{Acceleration}
\ee

The explicit values of the constants $\omega$ and $\kappa$ dictate the dominant accelerative behavior of the brane at initial as well as
large times. For example, larger values of $\kappa$ lead to more extreme early inflationary acceleration, while larger values of $\omega$ lead to slower, later time, accelerations. A model such as (\ref{Acceleration}) leads to
\be
    G_{i\bar j} \dot z^i \dot z^{\bar j}=\frac{{6(G_1  + 1)}}{{\omega ^{\rm 2} }}\left( {\frac{{e^{{t \mathord{\left/
         {\vphantom {t \omega }} \right.
         \kern-\nulldelimiterspace} \omega }}  + \omega \kappa e^{ - \kappa t} }}{{e^{{t \mathord{\left/
             {\vphantom {t \omega }} \right.
         \kern-\nulldelimiterspace} \omega }}  - e^{ - \kappa t} }}} \right)^2\label{ModuliDecay}
\ee
which is initially very large, then converges to a constant quantity:
\be
    \mathop {\lim }\limits_{t \to \infty }   G_{i\bar j} \dot z^i \dot z^{\bar j}= \frac{{6(G_1  + 1)}}{{\omega ^{\rm 2} }}.
\ee

\begin{figure}[hp]
    \centering
    \includegraphics[scale=0.6]{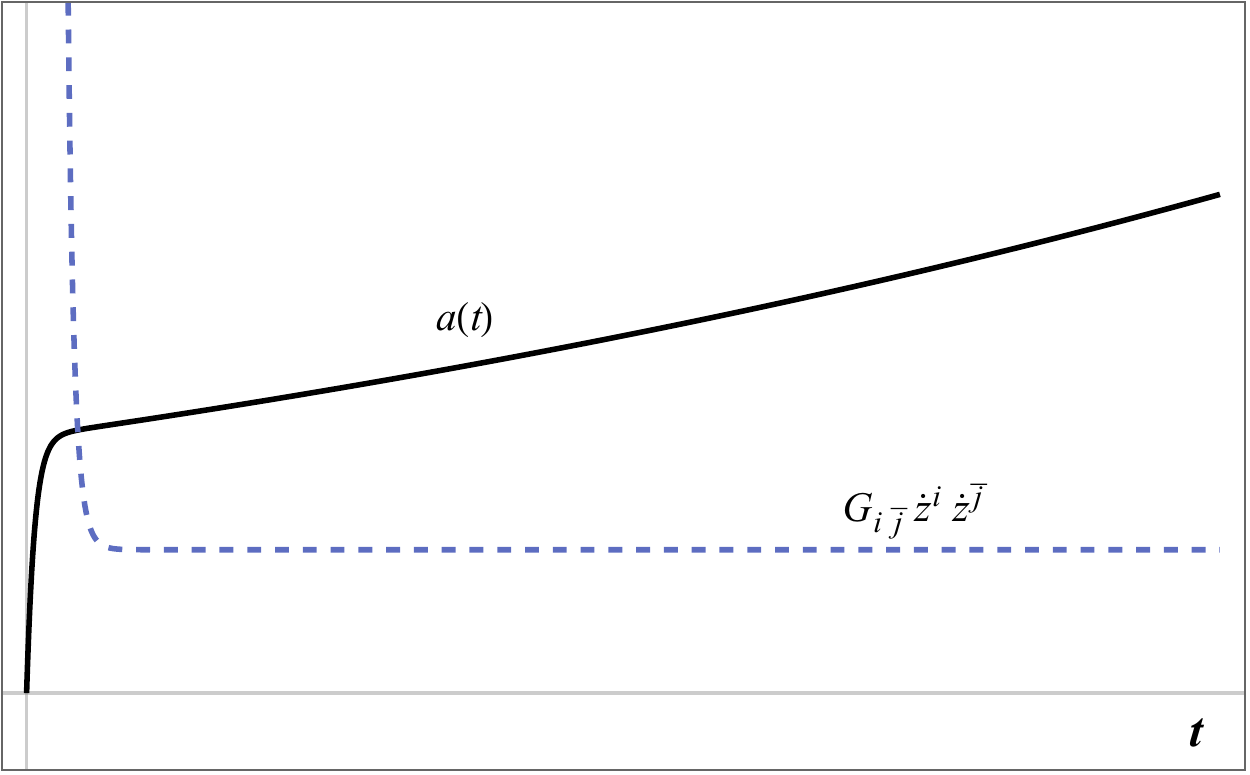}
    \caption{The correlation of the brane's scale factor (\ref{Acceleration}) with the norm of the moduli (\ref{ModuliDecay}).}
    \label{fig:1}
\end{figure}

Reversing the logic, this suggests that the complex structure moduli are highly unstable, decaying very rapidly at very early times, synchronous with an inflationary period. The decay rate is controlled by $\kappa$, which may then be thought of as related to the mass of the moduli. At later times, however, a slowly convergent value of the moduli coincides with a slow accelerative expansion. From that perspective, the value of $\omega$ can possibly be thought of as related to the original density of the moduli. This behavior is in perfect agreement with the prevalent understanding in the literature. Highly dense massive moduli produced in the very early hot universe and rapidly decaying to an almost constant value correlate with the accelerative expansion of the universe. Whether this correlation is a direct causality or just a bi-product of a more complex mechanism is an important question best left to future study. Fig (\ref{fig:1}) gives a comparative sketch of $a\left( t \right)$ and $G_{i\bar j} \dot z^i \dot z^{\bar j}$ based on this argument.

\pagebreak

\section{Conclusion}

We have constructed a multi 3-brane embedding in ungauged five dimensional $\N=2$ supergravity. We derived the dynamical equations of this spacetime as well as studied the general form of the hypermultiplet scalars in the bulk. A natural difficulty that arises in this and similar calculations is that the hypermultiplet fields are dependent on the unknown form of the underlying Calabi-Yau space. The best one can do is to find constraints on the fields, rather than explicit solutions. The symplectic structure of the theory was used to simplify these relations. An added difficulty is the complexity of the resulting constraints, particularly on the dilaton in this case. We did find one analytical solution for the dilaton's time dependence that unfortunately suffers from being too trivial. The other quantity that the hypermultiplet scalars depend on is an arbitrary function harmonic in the bulk dimension. Fortunately this function can be found analytically in terms of the metric warp functions.

We also showed that if one focuses on only one of the branes, its time evolution is dependent solely on the norm of the moduli of the complex structure of the Calabi-Yau. In cosmological terms, the moduli act as both a cosmological constant (or correction thereof) and inflaton potential. Focusing on specific samples of brane-universe evolution, the moduli exhibit what seems to be an instability. At very early times, their norm has a very large value that decays and converges to a constant quantity (which may be vanishing) at later times. In one particular case, the decay is very rapid and correlates directly with an early inflationary epoch.

In terms of a possible application of these results to our own universe's cosmological evolution, we note that the early production and decay of heavy moduli is required to explain the phenomenology of the early universe, in perfect agreement with our conclusions. There is still, however, a lot of work to be done. For example, the correlation between the brane's accelerative behavior and that of the decay of the moduli is largely unexplained. More in-depth analysis of the flow and evolution of moduli needs to be performed before a specific mechanism explaining this behavior can be pinpointed. In addition, the brane we studied was assumed vacuous, devoid of all matter or radiation. A more realistic model is needed before one can make a concrete connection with our own universe's cosmology. All of this we plan to explore in future work.

\pagebreak


\begin{thebibliography}{999}

\bibitem{Aharony:1999ti}
  O.~Aharony, S.~S.~Gubser, J.~M.~Maldacena, H.~Ooguri and Y.~Oz,
  ``Large N field theories, string theory and gravity,''
  Phys.\ Rept.\  {\bf 323}, 183 (2000)
  [hep-th/9905111].

\bibitem{Duff:1999rk}
  M.~J.~Duff,
  ``TASI lectures on branes, black holes and Anti-de Sitter space,''
  hep-th/9912164.

\bibitem{Randall:1999ee}
  L.~Randall and R.~Sundrum,
  ``A Large mass hierarchy from a small extra dimension,''
  Phys.\ Rev.\ Lett.\  {\bf 83}, 3370 (1999)
  [hep-ph/9905221].

\bibitem{Brax:2003fv}
  P.~Brax and C.~van de Bruck,
  ``Cosmology and brane worlds: A Review,''
  Class.\ Quant.\ Grav.\  {\bf 20}, R201 (2003)
  [hep-th/0303095].

\bibitem{Maartens:2010ar}
  R.~Maartens and K.~Koyama,
  ``Brane-World Gravity,''
  Living Rev.\ Rel.\  {\bf 13}, 5 (2010)
  [arXiv:1004.3962 [hep-th]].

\bibitem{Roane:2007zz}
  A.~F.~Roane,
  ``Some brane-world cosmological models,''
  AAT-3330715, PROQUEST-1650498761.

\bibitem{Flanagan:1999cu}
  E.~E.~Flanagan, S.~H.~H.~Tye and I.~Wasserman,
  ``Cosmological expansion in the Randall-Sundrum brane world scenario,''
  Phys.\ Rev.\ D {\bf 62}, 044039 (2000)
  [hep-ph/9910498].

\bibitem{Binetruy:1999hy}
  P.~Binetruy, C.~Deffayet, U.~Ellwanger and D.~Langlois,
  ``Brane cosmological evolution in a bulk with cosmological constant,''
  Phys.\ Lett.\ B {\bf 477}, 285 (2000)
  [hep-th/9910219].

\bibitem{Saaidi:2010jw}
  K.~Saaidi and A.~H.~Mohammadi,
  ``Brane Cosmology for Vacuum and Cosmological Constant Bulk,''
  arXiv:1006.1850 [gr-qc].

\bibitem{Saaidi:2012ri}
  K.~Saaidi and A.~Mohammadi,
  ``Brane Cosmology with the Chameleon Scalar Field in Bulk,''
  Phys.\ Rev.\ D {\bf 85}, 023526 (2012)
  [arXiv:1201.0371 [gr-qc]].

\bibitem{Lidsey:2000mt}
  J.~E.~Lidsey,
  ``Supergravity brane cosmologies,''
  Phys.\ Rev.\ D {\bf 62}, 083515 (2000)
  [hep-th/0007014].

\bibitem{Choudhury:2012ib}
  S.~Choudhury and S.~Pal,
  ``Brane inflation: A field theory approach in background supergravity,''
  J.\ Phys.\ Conf.\ Ser.\  {\bf 405}, 012009 (2012)
  [arXiv:1209.5883 [hep-th]].

\bibitem{Maia:2008yya}
  M.~D.~Maia,
  ``The Cosmological Constant Problem in Brane-world Cosmology,''
  EAS Publ.\ Ser.\  {\bf 30}, 319 (2008).

\bibitem{Okada:2014eva}
  N.~Okada and S.~Okada,
  ``Simple inflationary models in Gauss-Bonnet brane-world cosmology,''
  arXiv:1412.8466 [hep-ph].

\bibitem{Cordero:2011zz}
  R.~Cordero and E.~Rojas,
  ``Classical and quantum aspects of brane-world cosmology,''
  AIP Conf.\ Proc.\  {\bf 1396}, 55 (2011).

\bibitem{Capistrano:2011zz}
  A.~J.~S.~Capistrano and P.~I.~Odon,
  ``On the cosmological constant problem and brane-world geometry,''
  Central Eur.\ J.\ Phys.\  {\bf 9}, 189 (2011).

\bibitem{Amarilla:2009rs}
  L.~Amarilla and H.~Vucetich,
  ``Brane-world cosmology and varying G,''
  Int.\ J.\ Mod.\ Phys.\ A {\bf 25}, 3835 (2010)
  [arXiv:0908.2949 [gr-qc]].

\bibitem{Carmeli:1900zzc}
  M.~Carmeli,
  ``Cosmological general relativity in five dimensions: Brane world theory,''
  In *Carmeli, M. (ed.): Relativity: Modern large-scale spacetime structure of the cosmos* 233-263 (2008).

\bibitem{Koyama:2007rx}
  K.~Koyama,
  ``The cosmological constant and dark energy in braneworlds,''
  Gen.\ Rel.\ Grav.\  {\bf 40}, 421 (2008)
  [arXiv:0706.1557 [astro-ph]].

\bibitem{Antoniadis:2007hp}
  I.~Antoniadis, S.~Cotsakis and I.~Klaoudatou,
  ``Braneworld cosmological singularities,''
  gr-qc/0701033.

\bibitem{McFadden:2005mq}
  P.~L.~McFadden, N.~Turok and P.~J.~Steinhardt,
  ``Solution of a braneworld big crunch / big bang cosmology,''
  Phys.\ Rev.\ D {\bf 76}, 104038 (2007)
  [hep-th/0512123].

\bibitem{Garriga:2001qt}
  J.~Garriga,
  ``Brane world cosmology and the 5D big bang,''

\bibitem{Rasanen:2001hf}
  S.~Rasanen,
  ``On ekpyrotic brane collisions,''
  Nucl.\ Phys.\ B {\bf 626}, 183 (2002)
  [hep-th/0111279].

\bibitem{Khoury:2001wf}
  J.~Khoury, B.~A.~Ovrut, P.~J.~Steinhardt and N.~Turok,
  ``The Ekpyrotic universe: Colliding branes and the origin of the hot big bang,''
  Phys.\ Rev.\ D {\bf 64}, 123522 (2001)
  [hep-th/0103239].

\bibitem{Falkowski:2000er}
  A.~Falkowski, Z.~Lalak and S.~Pokorski,
  ``Supersymmetrizing branes with bulk in five-dimensional supergravity,''
  Phys.\ Lett.\ B {\bf 491}, 172 (2000)
  [hep-th/0004093].

\bibitem{astro-ph/9805201}
  A.~G.~Riess {\it et al.}  [Supernova Search Team Collaboration],
  ``Observational evidence from supernovae for an accelerating universe and a cosmological constant,''
  Astron.\ J.\  {\bf 116}, 1009 (1998)
  [astro-ph/9805201].

\bibitem{Linde:1994yf}
  A.~D.~Linde,
  ``Lectures on inflationary cosmology,''
  In *Rome 1994, Proceedings, Birth of the universe and fundamental physics* 363-372, and In *Lake Louise 1994, Proceedings, Particle physics and cosmology* 72-109, and Stanford U. - SU-ITP-94-36 (94,rec.Oct.) 44 p
  [hep-th/9410082].

\bibitem{AvilanV.:2010ri}
  N.~Avilan V. and J.~R.~Roldan,
  ``The Cosmological Constant Problem from the Point of View of String Theory,''
  arXiv:1011.5708 [hep-th].

\bibitem{1203.0307}
  R.~Bousso,
  ``The Cosmological Constant Problem, Dark Energy, and the Landscape of String Theory,''
  Pontif.\ Acad.\ Sci.\ Scr.\ Varia {\bf 119}, 129 (2011)
  [arXiv:1203.0307 [astro-ph.CO]].

\bibitem{Park:2013nu}
  E.~K.~Park and P.~S.~Kwon,
  ``Remark on Calabi-Yau vacua of the string theory and the cosmological constant problem,''
  Phys.\ Rev.\ D {\bf 88}, no. 4, 046007 (2013)
  [arXiv:1301.1783 [hep-th]].

\bibitem{Kallosh:2010xz}
  R.~Kallosh, A.~Linde and T.~Rube,
  ``General inflaton potentials in supergravity,''
  Phys.\ Rev.\ D {\bf 83}, 043507 (2011)
  [arXiv:1011.5945 [hep-th]].

\bibitem{Gong:2006be}
  J.~O.~Gong,
  ``On two string axions as inflaton,''
  hep-ph/0610423.

\bibitem{Hayashi:2014aua}
  H.~Hayashi, R.~Matsuda and T.~Watari,
  ``Issues in Complex Structure Moduli Inflation,''
  arXiv:1410.7522 [hep-th].

\bibitem{Emam:2010kt}
  M.~H.~Emam,
  ``The Many symmetries of Calabi-Yau compactifications,''
  Class.\ Quant.\ Grav.\  {\bf 27}, 163001 (2010)
  [arXiv:1007.4847 [hep-th]].

\bibitem{Emam:2009xj}
  M.~H.~Emam,
  ``Symplectic covariance of the N=2 hypermultiplets,''
  Phys.\ Rev.\ D {\bf 79}, 085017 (2009)
  [arXiv:0904.1951 [hep-th]].

\bibitem{Kallosh:2001du}
  R.~Kallosh, L.~Kofman, A.~D.~Linde and A.~A.~Tseytlin,
  ``BPS branes in cosmology,''
  Phys.\ Rev.\ D {\bf 64}, 123524 (2001)
  [hep-th/0106241].

\bibitem{Kabat:2001qt}
  D.~N.~Kabat and A.~Rajaraman,
  ``Testing cosmological supersymmetry breaking,''
  Phys.\ Lett.\ B {\bf 516}, 383 (2001)
  [hep-ph/0102309].

\bibitem{Canestaro:2013xsa}
  C.~A.~Canestaro and M.~H.~Emam,
  ``The five dimensional universal hypermultiplet and the cosmological constant problem,''
  Phys.\ Lett.\ B {\bf 726}, 913 (2013)
  [arXiv:1311.0266 [hep-th]].

\bibitem{Dine:2006ii}
  M.~Dine, R.~Kitano, A.~Morisse and Y.~Shirman,
  ``Moduli decays and gravitinos,''
  Phys.\ Rev.\ D {\bf 73}, 123518 (2006)
  [hep-ph/0604140].

\bibitem{Bodeker:2006ij}
  D.~Bodeker,
  ``Moduli decay in the hot early Universe,''
  JCAP {\bf 0606}, 027 (2006)
  [hep-ph/0605030].

\bibitem{Dudas:2012wi}
  E.~Dudas, A.~Linde, Y.~Mambrini, A.~Mustafayev and K.~A.~Olive,
  ``Strong moduli stabilization and phenomenology,''
  Eur.\ Phys.\ J.\ C {\bf 73}, no. 1, 2268 (2013)
  [arXiv:1209.0499 [hep-ph]].

\end{thebibliography}
\end{document}